\documentclass[sigconf]{acmart}

\usepackage{subcaption}
\usepackage{colortbl}
\usepackage{mathtools}
\usepackage{multirow}
\usepackage{enumitem}
\usepackage{makecell}
\usepackage{cleveref}
\usepackage{balance}

\AtBeginDocument{%
  }

\copyrightyear{2026}
\acmYear{2026}
\setcopyright{cc}
\setcctype{by}
\acmConference[MM '26]{Proceedings of the 34th ACM International Conference on Multimedia}{November 10--14, 2026}{Rio de Janeiro, Brazil}
\acmBooktitle{Proceedings of the 34th ACM International Conference on Multimedia (MM '26), November 10--14, 2026, Rio de Janeiro, Brazil}
\acmDOI{10.1145/3767308.3835152}
\acmISBN{979-8-4007-2213-4/2026/11}

\begin{document}

\title{Towards Transfer-Efficient Multi-modal Sequential Recommendation with State Space Duality}

\author{Hao Fan}
\authornote{Both authors contributed equally to this research.}
\orcid{0009-0004-2603-9365}
\affiliation{%
  \department{College of Mathematics and Computer Science}
  \institution{Zhejiang A\&F University}
  \city{Hangzhou}
  \state{Zhejiang}
  \country{China}
}
\email{fanhao986486@stu.zafu.edu.cn}

\author{Qingyang Liu}
\authornotemark[1]
\authornote{Corresponding authors: Qingyang Liu, Hongjiu Liu, and Kai Fang.}
\orcid{0000-0003-0491-3248}
\affiliation{%
  \department{School of Information Science and Engineering}
  \institution{East China University of Science and Technology}
  \city{Shanghai}
  \country{China}
}
\email{qingyang.liu@ecust.edu.cn}

\author{Hongjiu Liu}
\authornotemark[2]
\orcid{0000-0001-8175-264X}
\affiliation{%
  \department{College of Mathematics and Computer Science}
  \institution{Zhejiang A\&F University}
  \city{Hangzhou}
  \state{Zhejiang}
  \country{China}
}
\email{joe\_hunter@zafu.edu.cn}

\author{Yanrong Hu}
\orcid{0000-0002-9826-5212}
\affiliation{%
  \department{College of Mathematics and Computer Science}
  \institution{Zhejiang A\&F University}
  \city{Hangzhou}
  \state{Zhejiang}
  \country{China}
}
\email{yanrong\_hu@zafu.edu.cn}

\author{Kai Fang}
\authornotemark[2]
\orcid{0000-0003-0419-1468}
\affiliation{%
  \department{College of Mathematics and Computer Science}
  \institution{Zhejiang A\&F University}
  \city{Hangzhou}
  \state{Zhejiang}
  \country{China}
}
\email{kaifang@zafu.edu.cn}
\newcommand{\modelname}{MMM4Rec}
\settopmatter{printacmref=true}
\begin{abstract}
Sequential Recommendation (SR) models infer user preferences from interaction histories. While transferable Multi-modal SR models outperform traditional ID-based approaches, existing methods struggle with slow fine-tuning convergence due to complex optimization requirements and negative transfer effects. We propose \textbf{MMM4Rec} (\underline{M}ulti-\underline{M}odal \underline{M}amba for Sequential \underline{Rec}ommendation), a novel Multi-modal SR framework that incorporates a dedicated algebraic constraint mechanism for efficient transfer learning. By combining State Space Duality (SSD)'s temporal decay properties with a globally-aware temporal modeling design, our model dynamically prioritizes key modality information, overcoming limitations of Transformer-based approaches. The framework implements a constrained two-stage process: (1) sequence-level cross-modal alignment via shared projection matrices, followed by (2) temporal fusion using our newly designed Cross-SSD module and dual-channel Fourier adaptive filtering. This architecture maintains semantic consistency while suppressing noise propagation. By incorporating algebraic structural constraints aligned with SR priors, MMM4Rec employs a simple and consistent cross-entropy objective across both pre-training and fine-tuning, enabling rapid fine-tuning convergence, substantially improving multimodal recommendation accuracy, and preserving strong transferability. Extensive experiments demonstrate MMM4Rec's state-of-the-art performance, achieving strong multi-modal retrieval capability and exhibiting 10$\times$ faster average convergence speed when transferring to large-scale downstream datasets. The implementation is available at link \url{https://github.com/AlwaysFHao/MMM4Rec}.
\end{abstract}



\maketitle

\section{Introduction}
Recommender Systems (RS) are essential to platforms such as e-commerce and social media \cite{RSSurvey2015, LightGCN2020}. Sequential Recommendation (SR), an important RS subfield, learns user interests from interaction sequences to predict the next item \cite{SRS2019, SRSurvey2020}.
\par
Previous Sequential Recommenders predominantly rely on ID-based features \cite{SASRec2018, BERT4Rec2019, Mamba4Rec2024, TiM4Rec2024}. Despite their success, they have two inherent limitations: a) reliance on interaction data makes representation learning vulnerable to data sparsity and item cold-start \cite{ColdStart2002}; b) platform- and domain-specific ID mappings create inconsistent semantic spaces, hindering transfer and collaborative optimization across related domains \cite{MISSRec2023}. 
\begin{figure*}[t]
    \centering
    \includegraphics[width=0.9\textwidth]{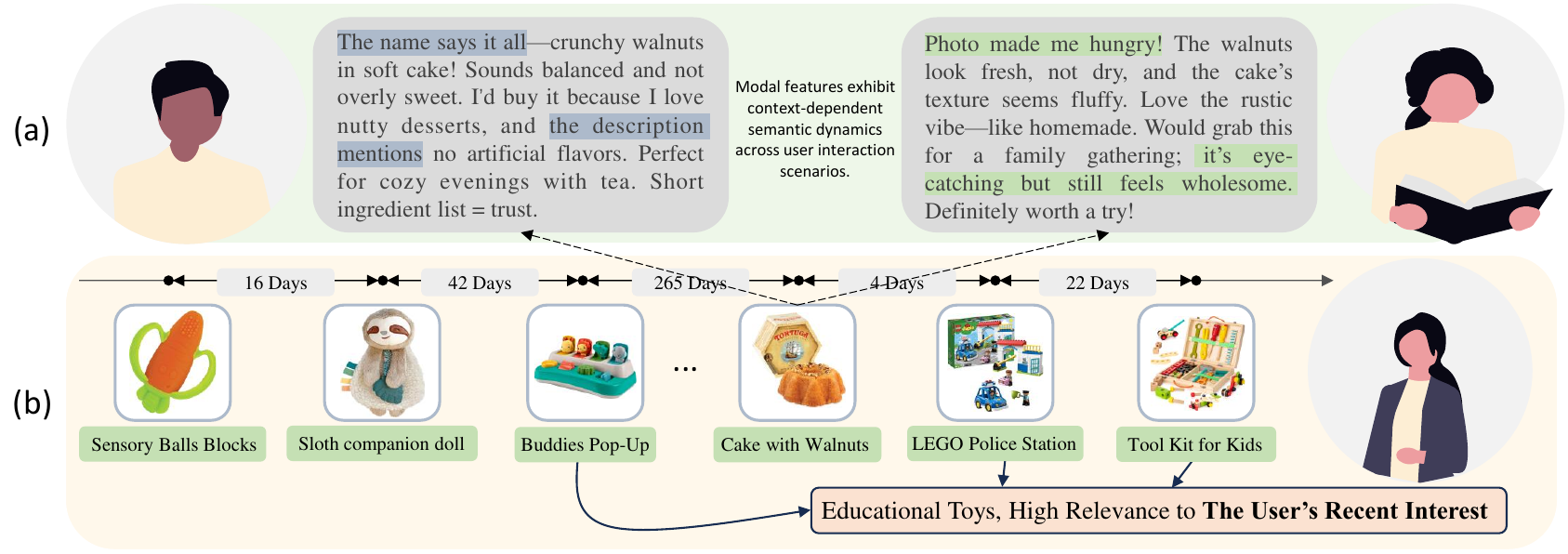} 
    \caption[The overview of \modelname{} .]{The two main research problems in multi-modal SR: (a) The alignment problem between multi-modal information and recommendation semantic space. (b) The unequal contribution problem of items within interaction sequences.}
    \label{fig:Example}
\end{figure*}
\par
With the remarkable advancements in computer vision (CV) \cite{VisonSurvey2022} and natural language processing (NLP) \cite{NLPSurvey2023}, researchers have identified the necessity and feasibility of introducing such modalities with general semantic representations into the field of SR to address the inherent limitations of ID-based models \cite{UniSRec2022, VQRec2023, MISSRec2023, PMMRec2024}. However, effectively utilizing multi-modal information in SR remains a significant challenge. Existing studies indicate that aligning the multi-modal semantic space with the recommendation semantic space is a critical factor in leveraging multi-modal information in SR \cite{VQRec2023, PMMRec2024}. While the solution to this problem is not yet fully understood, an effective approach involves pretraining the model on large-scale recommendation datasets using the generality of multi-modal semantic information \cite{TransRec2024, VQRec2023}. This imparts the model with prior knowledge of multi-modal information aligned with the recommendation semantic space, which can subsequently be fine-tuned on downstream datasets via transfer learning. To mitigate issues such as negative transfer \cite{VQRec2023} and the seesaw phenomenon \cite{SeesawPhenomenon2020}, as well as to guide the learning of effective multi-modal priors, existing research often employs complex contrastive learning strategies and cumbersome optimization processes to constrain the model's learning trajectory \cite{UniSRec2022, MISSRec2023, PMMRec2024}. However, these manually designed, non-end-to-end learning paradigms hinder the model's ability to achieve rapid convergence on downstream tasks. 
\textbf{This study investigates intrinsic algebraic constraints aligned with SR principles to help software engineers rapidly adapt pre-trained multi-modal recommenders to downstream tasks without intricate optimization objectives and procedures, enabling efficient knowledge transfer. }
\par
In practical sequential recommendation scenarios, effectively leveraging multi-modal information from user interaction sequences presents the following challenges (\cref{fig:Example}). 
\textbf{(\romannumeral1) Representation Alignment. }
The motivations behind different users interacting with the same item across varying temporal contexts are inherently diverse. This implies that static multi-modal features carry distinct semantic meanings under different interaction contexts, while the contribution weights of different modalities dynamically vary accordingly. Although multi-modal representation alignment serves as a common approach to address modality-specific contribution disparities, existing methods typically employ cross-modal contrastive learning strategies in recommendation semantic spaces \cite{MMSRec2023, PMMRec2024}. However, such approaches substantially increase model complexity and impede convergence speed, particularly due to the non-trivial task of designing appropriate negative sampling strategies tailored for recommendation semantics. While MISSRec \cite{MISSRec2023} achieves efficient alignment through user-specific modality fusion coefficients at the candidate item side, this method overlooks the learning process of sequence-level interest representations from the user perspective. A more optimal solution might lie in developing sequence-aware adaptive fusion mechanisms that collaboratively weigh modality contributions across varying interaction contexts.
\textbf{(\romannumeral2) Uneven Contribution Prioritization. }
Prior studies indicate that later-occurring items in interaction sequences generally better reflect users' current interest tendencies. Despite positional encoding enabling sequence ordering, transformer-based models initially treat multi-modal features of all items equally. This fundamental design might fail to prioritize modality information from recent items as theoretically expected. MISSRec addresses this through a multi-modal clustering approach to eliminate information redundancy and highlight critical item features. While effective, this clustering process breaks the end-to-end learning paradigm, and the suboptimal handcrafted feature modeling inevitably slows model convergence.
\par
To address these challenges, we introduce \textbf{\modelname{} } (\underline{M}ulti-\underline{M}odal \underline{M}amba for Sequential \underline{Rec}ommendation), a novel multi-modal framework designed for efficient and effective transferable learning in SR. Unlike conventional Transformer-based methods, our approach utilizes the state transition decay property of State Space Duality (SSD) \cite{SSD2024} and incorporates global temporal awareness to guide the prioritization of key modality information within user interaction sequences.
This design is also motivated by recent efficient SR backbones and multi-modal recommenders \cite{FMLPRec2023, LRURec2024, MM-Rec2022, CARCA2020, MMMLP2023}, which highlight the potential of combining efficient sequence modeling with richer item semantics.
In general, MMM4Rec takes interaction sequences with multi-modal information as input, learns to transform static multi-modal features into recommendation-aligned representations through simple pre-training, and achieves rapid downstream adaptation via specialized algebraic constraints.
Specifically, the proposed framework employs a two-stage multi-modal modeling process: alignment followed by fusion, both limited by algebraic constraints. In the alignment stage, cross-modal semantic alignment is achieved at the sequence level via a shared-parameter modal projection matrix, ensuring consistent multi-modal representations. During the fusion stage, we introduce a novel Cross-SSD module and a dual-channel Fourier-domain adaptive filter to capture temporal dependencies across modalities. These components enforce temporal consistency and correlation, maintaining semantic integrity while mitigating the influence of redundant or noisy information.
\par
The major contributions of this paper are:
\begin{itemize}[leftmargin=10pt]
    \item We develop a transferable multi-modal sequential recommender with dual advantages: multi-modal information effectiveness and fine-tuning efficiency.
    \item To effectively align multi-modal semantics with recommendation semantics, we propose an alignment-then-fusion approach for sequential modality integration, achieving robust multi-modal performance.
    \item By combining SSD's temporal decay with our temporal-aware enhancement, we develop efficient algebraic constraints for rapid capture of key modality patterns in user sequences.
    \item We propose a Time-aware Cross-SSD module dedicated to sequence-level multi-modal fusion.
    \item Through extensive experimentation covering both pre-training and diverse downstream fine-tuning scenarios, we provide conclusive evidence for \modelname{}'s effectiveness.
\end{itemize}
\section{Why Mamba Fits SR}
\label{sec:mamba_greater_than_transformer}
\begin{figure}[h]
    \centering
    \begin{subfigure}[b]{0.48\columnwidth}
        \centering
        \includegraphics[width=\linewidth]{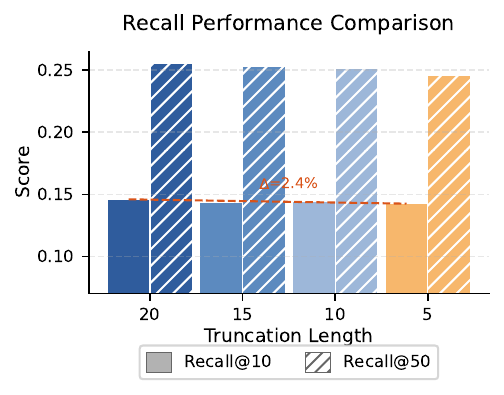}
    \end{subfigure}
    \hfill
    \begin{subfigure}[b]{0.48\columnwidth}
        \centering
        \includegraphics[width=\linewidth]{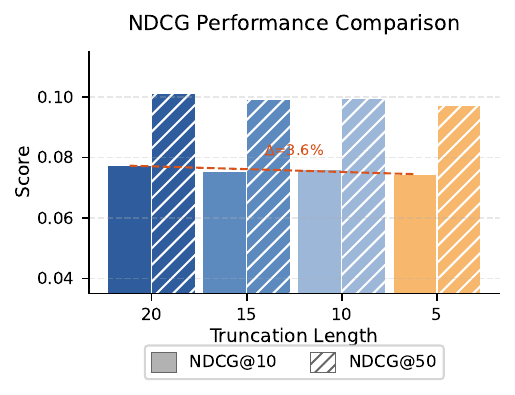}
    \end{subfigure}
    \caption{Performance of SASRec at different truncation lengths on the Kindle. }
    \label{fig:sasrec_truncation}
\end{figure}
\begin{figure*}[t]
    \centering
    \includegraphics[width=1\textwidth]{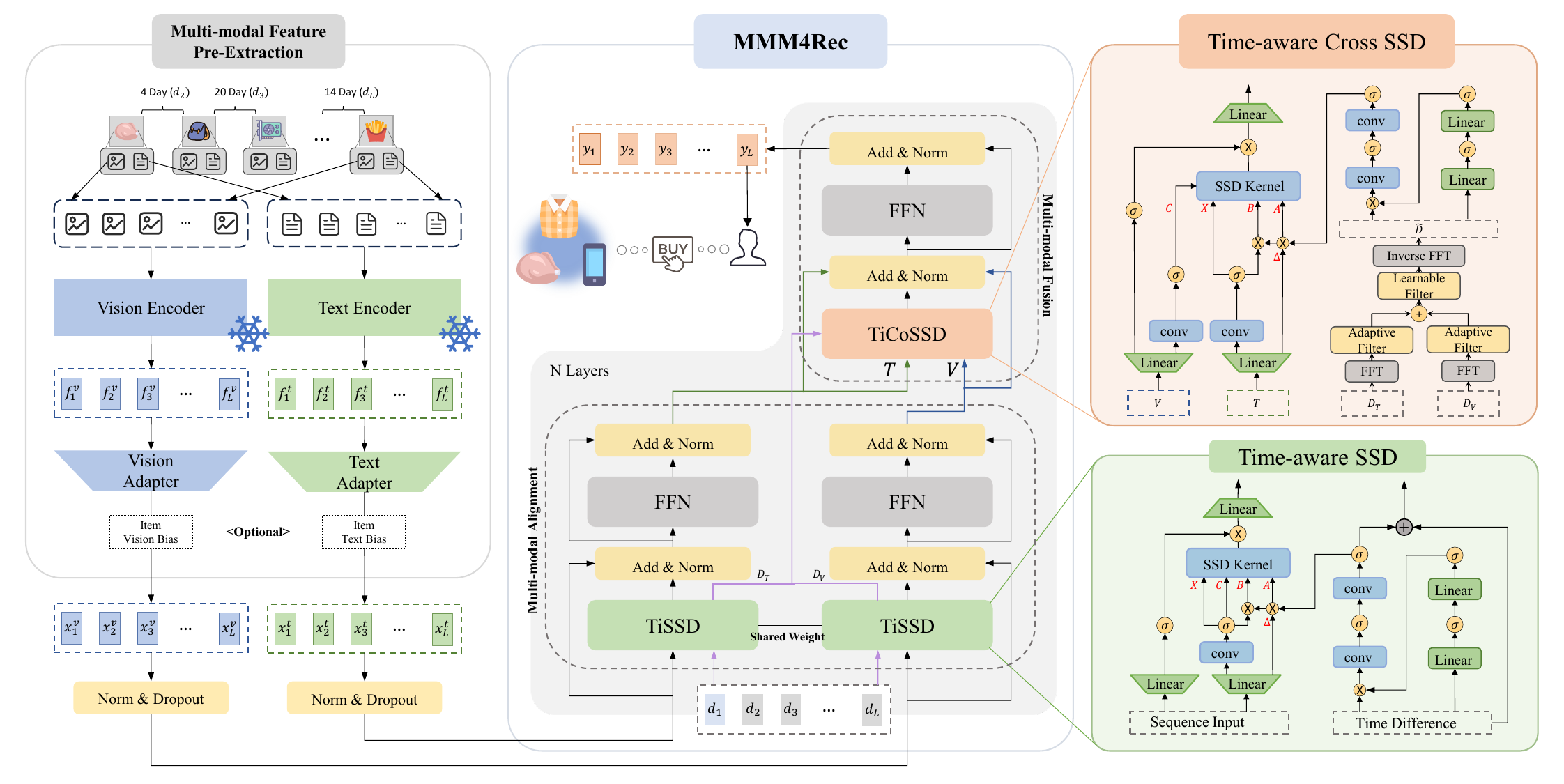} 
    \caption[The overview of \modelname{}.]{The overview of \modelname{}. }
    \label{fig:MMM4RecFramework}
\end{figure*}
Mamba4Rec's superior performance \cite{Mamba4Rec2024} over SASRec \cite{SASRec2018} at lower resource cost is consistent with Mamba's SR-aligned recency bias. SSD interprets Mamba as linear attention with a state-decay mask, naturally emphasizing recent interactions \cite{SSD2024,LinearAttention2020}. In our preliminary experiment on the Amazon \cite{Amazon2018} Kindle dataset, whose average interaction length is approximately 15, SASRec retains 97\% of the performance obtained with 20 items when using only the latest five items (\cref{fig:sasrec_truncation}). This result supports recency-aware modeling for faster downstream adaptation.

\section{Methods}
\subsection{Problem Formulation and Method Overview}
For user set $\mathcal{U}$ and item set $\mathcal{I}$, each user $u_{k} \in \mathcal{U}$ has a historical interaction sequence $\mathcal{S}^{u_k} = \left[ i_1, i_2, \cdots, i_L\right] \in \mathbb{R}^{L} $ (where $i_l \in \mathcal{I}$ denotes the $l$-th interacted item) ordered by interaction timestamps $\mathcal{T}^{u_k} = \left[t_1, t_2, \cdots, t_L\right] \in \mathbb{R}^{L}$, where $L$ is the number of interactions. The user/item population sizes are $|\mathcal{U}|$ and $|\mathcal{I}|$ respectively. In the multi-modal setting, every item $i \in \mathcal{I}$ is associated with unique image and text modal information $i^v$ and $i^t$. Sequential Recommender leverages historical interaction sequences to extract user interest representations, matches them with candidate items, and predicts the next item $i_{T + 1}$ that user $u_k$ is most likely to interact with. The overall architecture of MMM4Rec, as illustrated in \cref{fig:MMM4RecFramework}, comprises three core components: (\romannumeral1) multi-modal features are extracted through pre-trained frozen image/text encoders, followed by modality-specific adapters performing semantic transformation and dimensionality reduction. (\romannumeral2) Time-aware SSD with algebraic constraint implementation through inter-modality weight sharing achieves sequence-level cross-modal alignment. (\romannumeral3) A specially designed temporal-aware cross-SSD block fuses the aligned multi-modal information.
\par
These processes systematically address our target challenges through dual algebraic mechanisms: For cross-modal alignment, we implement sequence-level alignment via weight-sharing constraints that project different modalities into a unified recommendation space. For uneven item contributions, we exploit SSD's inherent algebraic constraint through its structured mask matrices that prioritize recent interactions (mathematically equivalent to emphasizing final sequence tokens), while augmenting this with our time-aware mask refinement - an algebraic extension modifying the original mask's eigenvalue distribution to preserve critical early interactions without compromising recent focus.

\subsection{Multi-modal Feature Pre-Extraction}
To obtain universal multi-modal representations of items, we employ an efficient multi-modal feature pre-extraction methodology.
\subsubsection{Pretrained Multi-modal Encoder}
We utilize cross-modally pretrained versions \cite{SigLIP2023} of BERT \cite{BERT4Rec2019} and ViT \cite{ViT2021} as the text modality encoder $\Phi^{t}$ and image modality encoder $\Phi^{v}$ respectively. We derive the user's text-modal feature sequence $\boldsymbol{F}^{t} = \left[f^{t}_{1}, f^{t}_{2}, \cdots, f^{t}_{L}\right] \in \mathbb{R}^{L \times D^{t}_{p}}$ and image-modal feature sequence $\boldsymbol{F}^{v} = \left[f^{v}_{1}, f^{v}_{2}, \cdots, f^{v}_{L}\right] \in \mathbb{R}^{L \times D^{v}_{p}}$ (where $D^{m}_{p}$ denotes the dimension of the modality-specific features extracted by the pretrained encoder corresponding to the $m$-th modality.) through the following transformation:
\begin{equation}
    \boldsymbol{F}^{v} = \Phi^{v} \left(\left[i^{v}_{1}, i^{v}_{2}, \cdots, i^{v}_{L}\right]\right), \quad \boldsymbol{F}^{t} = \Phi^{t} \left(\left[i^{t}_{1}, i^{t}_{2}, \cdots, i^{t}_{L}\right]\right). 
\end{equation}
\subsubsection{Modality-specific Adapters}
Aligned with MISSRec's parameter-efficient paradigm \cite{MISSRec2023}, we freeze the base parameters of pre-trained modality encoders and deploy lightweight modality-specific adapters \cite{Adapter1, Adapter2} for feature adaptation. This approach significantly reduces memory and computational overhead compared to full fine-tuning of the encoders, particularly when extracting multi-modal features across large-scale candidate item sets. Specifically, as formalized in \cref{eq:adapter}, the text-modal adapter $\Psi^{t}$ and image-modal adapter $\Psi^{v}$ transform raw modality features into rapidly adapted sequences $\boldsymbol{X}^t = \left[x^{t}_{1}, x^{t}_{2}, \cdots, x^{t}_{L}\right] \in \mathbb{R}^{L \times N}$ (text) and $\boldsymbol{X}^v = \left[x^{v}_{1}, x^{v}_{2}, \cdots, x^{v}_{L}\right] \in \mathbb{R}^{L \times N}$ (visual) respectively through constrained linear projections (where $N$ denotes the feature modeling dimension).
\begin{equation}
    \boldsymbol{X}^{v} = \Psi^{v} \left(\boldsymbol{F}^{v}\right) = \boldsymbol{F}^{v} W^{v}_{a} + b^{v}_{a}, \  \boldsymbol{X}^{t} = \Psi^{t} \left(\boldsymbol{F}^{v}\right) = \boldsymbol{F}^{t} W^{t}_{a} + b^{t}_{a}, 
    \label{eq:adapter}
\end{equation}
Where $W^{v}_{a} \in \mathbb{R}^{D^{v}_{p} \times N}$ and $W^{t}_{a} \in \mathbb{R}^{D^{t}_{p} \times N}$ represent the weight matrices,  while $b^{v}_{a}, b^{t}_{a} \in \mathbb{R}^{N}$  denote the corresponding bias vectors. 
\subsubsection{Optional Item Modality Bias}
\label{sec:modality-bias}
While sharing superficial preprocessing similarities with MISSRec \cite{MISSRec2023}, our architectural focus on \textbf{sequence-level multi-modal fusion} fundamentally distinguishes this work by rejecting early fusion of $\boldsymbol{X}^t$ and $\boldsymbol{X}^v$. To accelerate convergence in transfer learning, we propose a pluggable \textit{modality-gated item bias} module that injects domain-specific semantic priors (e.g., item popularity) via two trainable bias matrices $\mathbf{E}^t \in \mathbb{R}^{|\mathcal{I}| \times N}$ and $\mathbf{E}^v \in \mathbb{R}^{|\mathcal{I}| \times N}$. These matrices undergo element-wise addition to their corresponding modal features during inference, a mathematical formulation equivalent to learning static ID embeddings while bypassing the dimensionality explosion of explicit ID features.
Notably, these learned modality bias matrices are domain-specific, so they are discarded and re-learned when transferring to a new domain.
\subsection{Sequence-level Multi-modal Alignment}
We enable rapid convergence in sequence-level cross-modal recommendation semantic alignment through a carefully designed algebraic constraint mechanism compliant with sequential recommendation semantics. Specifically, the algebraic constraints consist of three components: (\romannumeral1) State Decay Constraint inherent to the State Space Duality (SSD) structure, which guides the model to prioritize the user's most recent interactions. (\romannumeral2) Temporal-aware Mask Matrix Constraint on SSD state transitions, preventing the model from neglecting critical early-interacted items. (\romannumeral3) Sequence-level Inter-modal Weight-Sharing Constraint that establishes intrinsic connections between modalities, enabling efficient collaborative optimization.
\subsubsection{Time-aware State Space Duality}
To enable efficient temporal-aware sequence modeling, we adopt the Time-aware SSD proposed in TiM4Rec \cite{TiM4Rec2024} for feature sequence extraction and semantic transformation. For an input sequence $\boldsymbol{X} \in \mathbb{R}^{L \times N}$, we generate variables $\boldsymbol{C}, \boldsymbol{B} \in \mathbb{R}^{L \times D}$ and $\Delta \in \mathbb{R}^{L}$ through the following transformations and process $\boldsymbol{X}$:
\begin{equation}
    [\boldsymbol{C}, \boldsymbol{B}, \boldsymbol{X}, \Delta] = \boldsymbol{X} W_1 + b_1, 
    \label{eq:ssd_input}
\end{equation}
\begin{equation*}
    W_1 \in \mathbb{R}^{N \times (2D + N + 1)}, b_1 \in \mathbb{R}^{2D + N + 1}. 
\end{equation*}
Subsequently, a causal convolution transformation \cite{Mamba2023} is applied to the matrices $ \boldsymbol{X} $, $ \boldsymbol{B} $ and $ \boldsymbol{C} $:
\begin{equation}
    \boldsymbol{X}_t, \boldsymbol{B}_t, \boldsymbol{C}_t = \sigma\left[(\boldsymbol{X}_t, \boldsymbol{B}_t, \boldsymbol{C}_t)^{\top} * \omega \right],
\end{equation}
{\small
\begin{equation*}
    \text{where} \quad \mathcal{Q}_t = \mathcal{P}_t * \omega \coloneqq \sum_{m=0}^{K-1} \mathcal{P}_{\max\left(t - m, 0\right)} \cdot \omega_{m}, 
\end{equation*}
}
Let $\omega \in \mathbb{R}^{K}$ denote the convolution kernel (kernel size K) and $\sigma\left(\cdot\right)$ the non-linear activation operator. 
\par
The state space discretization step size parameter $\Delta$ serves as the core parameter for generating SSD mask matrices. Crucially, the modeling granularity of $\Delta$ determines the specificity of SSD applications. By integrating the inter-item interaction time difference sequence $\mathcal{D} \in \mathbb{R}^{L}$ (See \cref{eq:time-diff}, where $LN$ denotes Layer Normalization \cite{LayerNorm2016}. ) into $\Delta$ through \cref{eq:hat_delta}, our model captures temporal patterns in user interaction behaviors, enabling explicit emphasis on critical items from interactions. 
\begin{equation}
    \mathcal{D} = LN\left(\left[0, \overline{d}_1, \overline{d}_2, \cdots, \overline{d}_{T - 1}\right]\right) = \left[d_0, d_1, d_2, \cdots, d_{T - 1}\right], \label{eq:time-diff}
\end{equation}
\begin{equation*}
    \overline{d_{l}} = t_{l + 1} - t_{l}, \quad t \in \mathcal{T}^{u_k},\quad l \in \left[1, T\right]
\end{equation*}
\begin{equation}
    \widehat{\mathcal{D}} = \alpha^{\mathcal{D}} \cdot \sigma\left(\mathcal{D} * \omega^{\mathcal{D}}\right), \quad \alpha^{\mathcal{D}} = MLP\left(\mathcal{D}\right), 
    \label{eq:time-enhance}
\end{equation}
\begin{equation}
    \hat{\Delta} = Softplus\left(\Delta \cdot \widehat{\mathcal{D}}\right) + b^{\Delta}, \quad b^{\Delta} \in \mathbb{R}^{L}. 
    \label{eq:hat_delta}
\end{equation}
The coefficient $\alpha^{\mathcal{D}}$ in \cref{eq:time-enhance} dynamically adjusts time differences using global user patterns, while causal convolution's local window enhances temporal pattern coverage.
\par
Following the Zero-Order Hold (ZOH) discretization scheme \cite{Mamba2023}, we discretize matrix $\boldsymbol{B}$ and the state space scalar coefficient $A \in \mathbb{R}^{1}$ in SSD \cite{SSD2024} using the time-aware augmented $\Delta$ through the following transformation:
\begin{equation}
    \overline{\boldsymbol{A}} = A \cdot \hat{\Delta}, \quad \overline{\boldsymbol{B}} = \hat{\Delta} \cdot \boldsymbol{B}, 
\end{equation}
Subsequently, we construct the Time-aware Structured Masked Matrix $\boldsymbol{L}$ as follows:
\begin{equation*}
    \hat{a}_i = \overline{\boldsymbol{A}}_i = A \cdot \hat{\Delta}_{i} = A \cdot \Delta_i \cdot d_i, 
\end{equation*}
\begin{equation}
    \boldsymbol{L} = \begin{bmatrix}
        \hat{a}_0 \\
        \hat{a}_1 & \hat{a}_0 \\
        \hat{a}_2 \hat{a}_1 & \hat{a}_2 & \hat{a}_0 \\
        \vdots & \vdots & \ddots & \ddots \\
        \hat{a}_{t - 1} \ldots  \hat{a}_1 & \hat{a}_{t - 1} \ldots \hat{a}_2 & \cdots & \hat{a}_{t - 1} & \hat{a}_0
    \end{bmatrix}, 
\end{equation}
Finally, the following equation can be derived to map the input sequence $\boldsymbol{X}$ and $\mathcal{D}$ to the output $\widetilde{\boldsymbol{X}} \in \mathbb{R}^{L \times N} $ and enhanced $\widehat{\mathcal{D}} \in \mathbb{R}^{L}$: 
\begin{equation}
    \widetilde{\boldsymbol{X}}, \widehat{\mathcal{D}} = TiSSD\left(\boldsymbol{X}, \mathcal{D}\right) \coloneqq \boldsymbol{L} \circ \boldsymbol{C} \overline{\boldsymbol{B}}^{\top} \boldsymbol{X}. 
    \label{eq:tissd}
\end{equation}
As analyzed Dao et al. \cite{SSD2024}, if matrix $\boldsymbol{C}$ is regarded as the query ( $\boldsymbol{Q}$ ) in attention mechanisms,  $\boldsymbol{\overline{B}}$ as keys ( $\boldsymbol{K}$ ) , and $\boldsymbol{X}$ as values ( $\boldsymbol{V}$ ), then SSD can be interpreted as a linear attention mechanism \cite{LinearAttention2020} with a specialized mask matrix. Leveraging the semi-separable block structure of matrix $\boldsymbol{L}$ and matrix associativity (by precomputing $\boldsymbol{K}^{\top}\boldsymbol{V}$), it achieves efficient linear attention computation with $O\left(T N^{2}\right)$ complexity. However, in multi-modal SR tasks where feature dimensions $N$ are typically large, traditional attention with $O\left(T^{2} N\right)$ complexity often dominates. To address this, we implement a mathematically equivalent squared attention formulation (TiSSD kernel), enabling flexible selection of the optimal SSD variant based on specific task dimensions. 

\subsubsection{Modal alignment of SR semantics}
For the image modality input feature sequence $\boldsymbol{X}^{v}$ and text modality input feature sequence $\boldsymbol{X}^{t}$, we implement weight-shared \cite{VLMo2022} constraint TiSSD to achieve efficient sequence-level cross-modal alignment compliant with SR semantics: 
\begin{equation}
    \begin{aligned}
        \widetilde{\boldsymbol{X}}^{v}, \widehat{\mathcal{D}}^{v} &= TiSSD\left(\boldsymbol{X^{v}}, \mathcal{D}\right), \boldsymbol{H}^{v} = LN\left(\widetilde{\boldsymbol{X}}^{v} + \boldsymbol{X}^{v}\right), \\
        \widetilde{\boldsymbol{X}}^{t}, \widehat{\mathcal{D}}^{t} &= TiSSD\left(\boldsymbol{X^{t}}, \mathcal{D}\right),  \boldsymbol{H}^{t} = LN\left(\widetilde{\boldsymbol{X}}^{t} + \boldsymbol{X}^{t}\right), \\
    \end{aligned}
    \label{eq:mm-tissd}
\end{equation}
\begin{equation}
    \begin{aligned}
        \boldsymbol{P}^{v} &= LN \left(FFN_{v}\left(\boldsymbol{H}^{v}\right) + \boldsymbol{H}^{v}\right), \\
        \boldsymbol{P}^{t} &= LN \left(FFN_{t}\left(\boldsymbol{H}^{t}\right) + \boldsymbol{H}^{t}\right), 
    \end{aligned}
\end{equation}
where the $LN$ denotes Layer Normalization \cite{LayerNorm2016} and the $FFN$ refers to Feed Forward Network that is consistent with the definition in Transformer \cite{Transformer2017}. 
The TiSSD modules for both modalities in \cref{eq:mm-tissd} are weight-shared. This sequence-level constraint compels the feature sequence extraction results of image and text modalities to be projected into a convergent recommendation semantic space. Through the aforementioned modality-specific feature extraction and transformation, we obtain semantically aligned image-modality feature sequence $\boldsymbol{P}^{v}$ and text-modality feature sequence $\boldsymbol{P}^{t}$ under the SR semantics. 

\subsection{Sequential-level Multi-modal Fusion}
After obtaining semantically aligned image and text modality feature sequences, we fuse these cross-modal sequences to derive unified user interest representations. To this end, we propose a novel Time-aware Cross SSD (TiCoSSD) module that achieves effective sequence-level multi-modal fusion. Specifically, TiCoSSD introduces two critical enhancements compared to TiSSD:
(\romannumeral1)Dual-Channel Fourier Filtering: Designed to integrate temporal patterns from both modalities through parallel frequency-domain transformations.
(\romannumeral2)Cross-Attention Inspired Structural Adaptation: By drawing inspiration from cross-attention mechanisms, we reconfigure the original SSD architecture to enable robust fusion of multi-modal feature sequences.
\subsubsection{Dual-Channel Fourier Filtering}
To capture user interaction temporal patterns suitable for the multi-modal fusion phase, we perform frequency-domain fusion on the time difference signals of both modalities. Specifically, for the time difference vectors $\widehat{\mathcal{D}}^{v}$ and $\widehat{\mathcal{D}}^{t}$ (refer to \cref{eq:mm-tissd}) output by the multi-modal alignment phase, we apply Fast Fourier Transform (FFT) as follows:
\begin{equation}
    \widetilde{\mathcal{D}}^{v} = \mathcal{F}\left(\widehat{\mathcal{D}}^{v}\right) \in \mathbb{C}^{L}, \quad \widetilde{\mathcal{D}}^{t} = \mathcal{F}\left(\widehat{\mathcal{D}}^{t}\right) \in \mathbb{C}^{L}, 
\end{equation}
where $\mathcal{F}\left(\cdot\right)$ denotes the 1-D FFT, and $\widetilde{\mathcal{D}}^{v}$ and $\widetilde{\mathcal{D}}^{t}$ are complex-valued spectra of the two modalities. We further decompose the filtering process into two modules.
\par
\textbf{Adaptive Filter.}
The adaptive filter generates modality-specific frequency kernels from each spectrum and applies element-wise filtering:
\begin{equation*}
    \widetilde{\delta}\left(\widetilde{\mathcal{D}}\right) \coloneqq \widetilde{\mathcal{D}} \widetilde{W} + \widetilde{b}, \quad \widetilde{W} \in \mathbb{C}^{L \times L}, \widetilde{b} \in \mathbb{C}^{L}
\end{equation*}
\begin{equation}
    \boldsymbol{K}^{v} = \widetilde{\delta}\left(\widetilde{\mathcal{D}}^{v}\right) \in \mathbb{C}^{L}, \quad \widetilde{\mathcal{D}}^{v}_{\text{filtered}} = \boldsymbol{K}^{v} \odot \widetilde{\mathcal{D}}^{v},
    \label{eq:adaptive-filter-v}
\end{equation}
\begin{equation}
    \boldsymbol{K}^{t} = \widetilde{\delta}\left(\widetilde{\mathcal{D}}^{t}\right) \in \mathbb{C}^{L}, \quad \widetilde{\mathcal{D}}^{t}_{\text{filtered}} = \boldsymbol{K}^{t} \odot \widetilde{\mathcal{D}}^{t},
    \label{eq:adaptive-filter-t}
\end{equation}
Here, $\boldsymbol{K}^{v}$ and $\boldsymbol{K}^{t}$ adaptively reweight frequency components for each modality. We clarify that $\widetilde{\delta}\left(\cdot\right)$ is computationally realized in PyTorch \cite{Pytorch2019} via:
\begin{equation}
    \begin{bmatrix}
        \Re(\boldsymbol{K}) \\
        \Im(\boldsymbol{K})
    \end{bmatrix}
    =
    \begin{bmatrix}
        \Re(\widetilde{W}) & -\Im(\widetilde{W}) \\
        \Im(\widetilde{W}) & \Re(\widetilde{W})
    \end{bmatrix}
    \begin{bmatrix}
        \Re(\widetilde{\mathcal{D}}) \\
        \Im(\widetilde{\mathcal{D}})
    \end{bmatrix}
    +
    \begin{bmatrix}
        \Re(\widetilde{b}) \\
        \Im(\widetilde{b})
    \end{bmatrix}.
\end{equation}
\textbf{Learnable Filter.}
After adaptive filtering, we fuse the two filtered spectra and apply another complex-valued linear transform before projecting back to the time domain:
\begin{equation}
    \widehat{\mathcal{D}}^{f} = \mathcal{F}^{-1}\left(\widetilde{\delta}\left(\widetilde{\mathcal{D}}^{v}_{\text{filtered}} + \widetilde{\mathcal{D}}^{t}_{\text{filtered}}\right)\right) \in \mathbb{C}^{L}, 
    \label{eq:mm-time}
\end{equation}
where $\mathcal{F}^{-1}(\cdot)$ denotes the inverse 1D FFT. This learnable transform refines the fused frequency representation and yields a unified time-difference signal for subsequent multi-modal fusion.
\subsubsection{Time-aware Cross SSD}
To fuse information from both modalities, we structurally adapt the original TiSSD by decoupling matrices $\boldsymbol{C}$, $\boldsymbol{B}$, and $\boldsymbol{X}$ through cross-attention \cite{Transformer2017} inspired operations. Specifically, we reformulate \cref{eq:ssd_input} in TiSSD as follows:
\begin{equation}
    \begin{aligned}
        \boldsymbol{C} &= \boldsymbol{P}^{v} W_2 + b_2, \  W_2 \in \mathbb{R}^{N \times D}, b_2 \in \mathbb{R}^{D}, \\
        [\boldsymbol{B}, \boldsymbol{X}, \Delta] &= \boldsymbol{P}^{t} W_3 + b_3, \  W_3 \in \mathbb{R}^{N \times \left(M\right)}, b_3 \in \mathbb{R}^{\left(M\right)}. 
    \end{aligned}
\end{equation}
Where $\left(M\right) = D + N + 1$. The final fused sequence $\boldsymbol{Y} \in \mathbb{R}^{L \times N}$ is derived by replacing $\mathcal{D}$ in \cref{eq:time-diff} with the cross-modal representation $\widehat{\mathcal{D}}^{f}$: 
By substituting the time difference parameter $\mathcal{D}$ in \cref{eq:time-diff} with the cross-modal representation $\widehat{\mathcal{D}}^{f}$ derived from \cref{eq:mm-time}, while retaining all other computational components from TiSSD, we obtain the multi-modally fused feature sequence $\boldsymbol{M} \in \mathbb{R}^{L \times N}$:
\begin{equation}
        \boldsymbol{M} = TiCoSSD\left(\boldsymbol{P}^{v},\boldsymbol{P}^{t}, \widehat{\mathcal{D}}^{f}\right) \coloneqq \boldsymbol{L} \circ \boldsymbol{C} \boldsymbol{B}^{\top} \left(\hat{\Delta}^{\top} \boldsymbol{X}\right). 
\end{equation}
Finally, we apply the following fundamental transformation to derive the final user interest representation sequence $\boldsymbol{Y} \in \mathbb{R}^{L \times N}$:
\begin{equation}
        \begin{aligned}
            \boldsymbol{O} &= LN\left(\boldsymbol{M} + \boldsymbol{P}^{v} + \boldsymbol{P}^{t}\right) \in \mathbb{R}^{L \times N}, \\
            \boldsymbol{Y} &= LN\left(FFN\left(\boldsymbol{O}\right) + \boldsymbol{O}\right). 
        \end{aligned}
\end{equation}

\subsection{Efficient Transfer Training Strategy}
To avoid convergence bottlenecks from complex objectives, we adopt a minimalist transfer learning strategy guided by Occam's razor.
\subsubsection{Multi-modal Candidate Item Score Calculation}
We use the last hidden state $y_L \in \mathbb{R}^{1 \times N}$ as the current user-interest representation $u_k$, and compute its score with candidate item $i_m$ by:
\begin{equation}
    \langle u_k, i_m \rangle = u_k \left[\Psi^{v}\left(\Phi^{v} \left(i_{m}^{v}\right)\right)\right]^{\top} + u_k \left(\Psi^{t}\left[\Phi^{t} \left(i_{m}^{t}\right)\right)\right]^{\top}, 
\end{equation}
where $i_{m}^{v}$ and $i_{m}^{t}$ denote the raw image and text of $i_m$. In practice, candidate features are pre-extracted by the pretrained modal encoder $\Phi\left(\cdot\right)$ offline.
\subsubsection{Minimalist Pre-training}
The pre-training phase uses only standard cross-entropy loss, without auxiliary objectives. Given the scale of pre-training data, we adopt in-batch negative sampling instead of full-corpus ranking to improve efficiency. The objective \emph{w.r.t.} $u_k$ is:
\begin{equation}
    \ell_{u_k}^{pre-train} = -\log \frac{\exp \left(\langle u_k, i_{L_{u_k} + 1} \rangle / \tau \right)}{\sum_{j=1}^{B} \exp \left(\langle u_j, i_{L_{u_j} + 1}\rangle / \tau \right)}, 
    \label{eq:loss-pretrain}
\end{equation}
where $B$ is the mini-batch size and $\tau > 0$ is the temperature.
\subsubsection{Fine-tuning}
Fine-tuning also uses standard cross-entropy loss. Since downstream data is much smaller, we switch to full-corpus ranking for negative sampling. The objective \emph{w.r.t.} $u_k$ is:
\begin{equation}
    \ell_{u_k}^{fine-tune} = -\log \frac{\exp \left(\langle u_k, i_{L_{u_k} + 1} \rangle / \tau \right)}{\sum_{j=1}^{|\mathcal{I}|} \exp \left(\langle u_j, i_j\rangle / \tau \right)}, 
    \label{eq:loss-funtine}
\end{equation}
\section{Experiments}
We evaluate the proposed method through pre-training on five datasets and conducting transfer learning on five downstream domain datasets. Our study addresses the following research questions:
\begin{itemize}
    \item[\textbf{RQ1: }] Compared to state-of-the-art (SOTA) SR models that explicitly utilize heterogeneous information, does \modelname{} achieve competitive performance in downstream domains?
    \item[\textbf{RQ2: }] Can \modelname{} achieve more transfer-efficient convergence when applied to downstream tasks?
    \item[\textbf{RQ3: }] How do different designs contribute to \modelname{}'s efficacy?
\end{itemize}
\subsection{Experimental Setup}
\subsubsection{Datasets}
We use 10 domains from \textbf{Amazon Reviews} \cite{Amazon2018}: \emph{Grocery and Gourmet Food}, \emph{Home and Kitchen}, \emph{CDs and Vinyl}, \emph{Kindle Store}, \emph{Movies and TV}, \emph{Prime Pantry}, \emph{Industrial and Scientific}, \emph{Musical Instruments}, \emph{Arts, Crafts and Sewing}, and \emph{Office Products}. The first five serve as pre-training domains and the latter five as downstream targets. Following \cite{UniSRec2022, MISSRec2023}, we apply 5-core filtering, extract textual metadata (titles, categories, and brands), and download product images from the provided URLs. As shown in \cref{tab:dataset}, text is complete, but many items lack images due to expired URLs. Following \cite{MISSRec2023}, we retain modality-missing items for fair comparison.
\begin{table}[ht] %
	\caption{Statistics of Pre-processed Datasets. ``Cover.'' denotes the image coverage among the item set. ``Avg. SL'' denotes the average length of interaction sequences.}
	\label{tab:dataset}
	\resizebox{\columnwidth}{!}{
	\begin{tabular}{l *{5}{r}}
		\toprule
		Datasets & \#Users & \#Items & \#Img. (Cover./\%) & \#Inters. & Avg. SL.\\
            \hline
		\midrule
		\emph{Pre-trained} & 1,361,408 & 446,975 & 94,151 (21.06\%) & 14,029,229 & 13.51 \\
		- Food   & 115,349 &  39,670 & 29,990 (75.60\%) & 1,027,413 &  8.91 \\
		- CDs    &  94,010 &  64,439 & 21,166 (32.85\%) & 1,118,563 & 12.64 \\
		- Kindle & 138,436 &  98,111 & 0 (0\%) & 2,204,596 & 15.93 \\
		- Movies & 281,700 &  59.203 & 8,675 (14.65\%) & 3,226,731 & 11.45 \\
		- Home   & 731,913 & 185,552 & 34,320 (18.50\%) & 6,451,926 &  8.82 \\
		\midrule
		Scientific  &  8,442 &  4,385 & 1,585 (36.15\%) &  59,427 & 7.04 \\
		Pantry      & 13,101 &  4,898 & 4,587 (93.65\%) & 126,962 & 9.69 \\
		Instruments & 24,962 &  9,964 & 6,289 (63.12\%) & 208,926 & 8.37 \\
		Arts        & 45,486 & 21,019 & 9,437 (44.90\%) & 395,150 & 8.69 \\
		Office      & 87,436 & 25,986 & 16,628 (63.99\%) & 684,837 & 7.84 \\ 
		\bottomrule
	\end{tabular}
	}
\end{table}
\subsubsection{Metrics}
Following \cite{UniSRec2022, MISSRec2023}, we evaluate retrieval performance with Recall@K (R$@K$) and NDCG@K (N$@K$). For a more comprehensive evaluation, we report results at $K \in \{10, 50\}$.
\subsubsection{Baselines}
We compare with 14 SOTA sequential recommenders: \textbf{(\romannumeral1)} ID-based models: SASRec \cite{SASRec2018}, Mamba4Rec \cite{Mamba4Rec2024}, TiM4Rec \cite{TiM4Rec2024}, and BSARec \cite{BSARec2024}; \textbf{(\romannumeral2)} text-enhanced models: ZESRec \cite{ZESRec2021}, FDSA \cite{FDSA2019}, S$^3$-Rec \cite{S3-Rec2020}, UniSRec \cite{UniSRec2022}, and VQRec \cite{VQRec2023}; \textbf{(\romannumeral3)} multi-modal models: MMSRec \cite{MMSRec2023}, MISSRec \cite{MISSRec2023}, M$^3$Rec \cite{M3Rec2025}, HM4SR \cite{HM4SR2025}, and ATHWE \cite{ATHWE2026}. We also derive text-enhanced variants from the first three ID-based models. Following \cite{UniSRec2022}, we use the official S$^3$-Rec setting for consistent representation learning. TiM4Rec, HM4SR, and ATHWE are time-aware enhanced models. Mamba4Rec, TiM4Rec, and M$^3$Rec are Mamba-based, while BSARec combines attention with Fourier filtering. UniSRec, VQRec, MMSRec, and MISSRec are transferable recommenders.
\subsubsection{Implementation Details}
We optimize with NAdam \cite{NAdam2016} (learning rate 1e-4), pre-train for 40 epochs, and apply early stopping with patience 10 during fine-tuning. SigLip-B/16 \cite{SigLIP2023} is used as the feature encoder, with modality adapters projecting features into a 256-dimensional latent space. For the Mamba backbone, we set the SSM state factor to 64, the 1D causal convolution kernel size to 4, and the block expansion factor to 2. To address Amazon sparsity \cite{Amazon2018}, we use dropout 0.4 and set $\tau=0.8$ in \cref{eq:loss-pretrain,eq:loss-funtine}. TiSSD and TiCoSSD both use a single stacked layer. All baselines follow their optimal reported settings with necessary adjustments for fair comparison.
\begin{table*}[ht]
\centering
\caption{Comparisons on different target datasets. 
``T'' and ``V'' stands for text and visual features. 
``\emph{Improv}.'' denotes the statistically significant relative improvement of \modelname{} to the best baselines ($t$-test, $p$-value $<$ 0.05). 
The best and second-best results are in bold and underlined. }
\setlength{\tabcolsep}{0.5em}{
 \resizebox{1\textwidth}{!}{
\begin{tabular}{lclccccccccccccr}
\toprule
\multicolumn{2}{l}{\makecell{Input Type \& Model $\rightarrow$}} &  & \multicolumn{4}{c}{ID} & \multicolumn{3}{c}{T+ID} & \multicolumn{5}{c}{T+V+ID} & \multirow{2}{*}{\makecell{\emph{Improv.} \\ w/ ID}}\\
\cmidrule(l){1-2} \cmidrule(l){4-7} \cmidrule(l){8-10} \cmidrule(l){11-15}
Dataset & Metric &  & {\small SASRec} & {\footnotesize Mamba4Rec} & {\small TiM4Rec} & {\small BSARec} & FDSA & S$^3$-Rec & {\small UniSRec} & {\small MISSRec} & M$^3$Rec & {\footnotesize HM4SR} & {\footnotesize{ATHWE}} & {\small \modelname{}} & \\
\hline
\midrule
\multirow{4}{*}{Scientific} & R@10 &  & 0.1080 & 0.1040 & 0.1079 & 0.1102 & 0.0899 & 0.0525 & 0.1235 & \textbf{0.1360} & 0.1105 & 0.0937 & 0.1070 & \underline{0.1348} & \cellcolor[HTML]{E9EAFF}- \\
 & R@50 &  & 0.2042 & 0.2030 & 0.2021 & 0.2106 & 0.1732 & 0.1418 & \underline{0.2473} & 0.2431 & 0.2142 & 0.1686 & 0.2072 & \textbf{0.2627} & \cellcolor[HTML]{E9EAFF}6.23\% \\
 & N@10 &  & 0.0553 & 0.0598 & 0.0605 & 0.0605 & 0.0580 & 0.0275 & 0.0634 & \textbf{0.0753} & 0.0616 & 0.0651 & 0.0711 & \underline{0.0724} & \cellcolor[HTML]{E9EAFF}- \\
 & N@50 &  & 0.0760 & 0.0814 & 0.0810 & 0.0824 & 0.0759 & 0.0468 & 0.0904 & \underline{0.0983} & 0.0842 & 0.0814 & 0.0802 & \textbf{0.1002} & \cellcolor[HTML]{E9EAFF}1.93\% \\
 \midrule
\multirow{4}{*}{Pantry} & R@10 &  & 0.0501 & 0.0487 & 0.0504 & 0.0531 & 0.0395 & 0.0444 & 0.0693 & \underline{0.0779} & 0.0495 & 0.0437 & 0.0573 & \textbf{0.0984} & \cellcolor[HTML]{E9EAFF}26.32\% \\
 & R@50 &  & 0.1322 & 0.1377 & 0.1360 & 0.1408 & 0.1151 & 0.1315 & 0.1827 & \underline{0.1875} & 0.1407 & 0.1156 & 0.1414 & \textbf{0.2127} & \cellcolor[HTML]{E9EAFF}13.44\% \\
 & N@10 &  & 0.0218 & 0.0223 & 0.0229 & 0.0234 & 0.0209 & 0.0214 & 0.0311 & \underline{0.0365} & 0.0222 & 0.0232 & 0.0314 & \textbf{0.0481} & \cellcolor[HTML]{E9EAFF}31.78\% \\
 & N@50 &  & 0.0394 & 0.0415 & 0.0411 & 0.0423 & 0.0370 & 0.0400 & 0.0556 & \underline{0.0598} & 0.0418 & 0.0388 & 0.0494 & \textbf{0.0729} & \cellcolor[HTML]{E9EAFF}21.91\% \\
\midrule
\multirow{4}{*}{Instruments} & R@10 &  & 0.1118 & 0.1113 & 0.1113 & 0.1156 & 0.1070 & 0.1056 & 0.1267 & \underline{0.1300} & 0.1145 & 0.1079 & 0.1193 & \textbf{0.1330} & \cellcolor[HTML]{E9EAFF}2.31\% \\
 & R@50 &  & 0.2106 & 0.2034 & 0.2071 & 0.2114 & 0.1890 & 0.1927 & \underline{0.2387} & 0.2370 & 0.2114 & 0.1881 & 0.2088 & \textbf{0.2525} & \cellcolor[HTML]{E9EAFF}5.78\% \\
 & N@10 &  & 0.0612 & 0.0751 & 0.0683 & 0.0649 & 0.0796 & 0.0713 & 0.0748 & \underline{0.0843} & 0.0764 & 0.0807 & \textbf{0.0872} & 0.0822 & \cellcolor[HTML]{E9EAFF}- \\
 & N@50 &  & 0.0826 & 0.0950 & 0.0890 & 0.0857 & 0.0972 & 0.0901 & 0.0991 & \underline{0.1071} & 0.0975 & 0.0979 & 0.1066 & \textbf{0.1082} & \cellcolor[HTML]{E9EAFF}1.03\% \\
\midrule
\multirow{4}{*}{Arts} & R@10 &  & 0.1108 & 0.1089 & 0.1096 & 0.1105 & 0.1002 & 0.1003 & 0.1239 & \textbf{0.1314} & 0.1098 & 0.1011 & 0.1123 & \underline{0.1307} & \cellcolor[HTML]{E9EAFF}- \\
 & R@50 &  & 0.2030 & 0.2036 & 0.2027 & 0.2102 & 0.1779 & 0.1888 & 0.2347 & \underline{0.2410} & 0.2027 & 0.1745 & 0.2007 & \textbf{0.2486} & \cellcolor[HTML]{E9EAFF}3.15\% \\
 & N@10 &  & 0.0587 & 0.0628 & 0.0630 & 0.0660 & 0.0714 & 0.0601 & 0.0712 & 0.0767 & 0.0636 & 0.0715 & \underline{0.0769} & \textbf{0.0777} & \cellcolor[HTML]{E9EAFF}1.04\% \\
 & N@50 &  & 0.0788 & 0.0834 & 0.0832 & 0.0877 & 0.0883 & 0.0793 & 0.0955 & \underline{0.1002} & 0.0838 & 0.0874 & 0.0971 & \textbf{0.1034} & \cellcolor[HTML]{E9EAFF}3.19\% \\
\midrule
\multirow{4}{*}{Office} & R@10 &  & 0.1056 & 0.1234 & 0.1227 & 0.1194 & 0.1118 & 0.1030 & \underline{0.1280} & 0.1275 & 0.1217 & 0.1142 & 0.1223 & \textbf{0.1337} & \cellcolor[HTML]{E9EAFF}4.45\% \\
 & R@50 &  & 0.1627 & 0.1886 & 0.1892 & 0.1878 & 0.1665 & 0.1613 & \underline{0.2016} & 0.2005 & 0.1864 & 0.1664 & 0.1797 & \textbf{0.2132} & \cellcolor[HTML]{E9EAFF}5.75\% \\
 & N@10 &  & 0.0710 & 0.0874 & 0.0876 & 0.0817 & 0.0868 & 0.0653 & 0.0831 & 0.0856 & 0.0858 & 0.0887 & \textbf{0.0947} & \underline{0.0906} & \cellcolor[HTML]{E9EAFF}- \\
 & N@50 &  & 0.0835 & 0.1016 & 0.1021 & 0.0966 & 0.0987 & 0.0780 & 0.0991 & 0.1012 & 0.0999 & 0.1001 & \underline{0.1071} & \textbf{0.1080} & \cellcolor[HTML]{E9EAFF} 0.84\% \\
\hline
\bottomrule
\end{tabular}
}
}
\label{tab:main_results}
\end{table*}
\begin{table*}[ht]
\centering
\caption{Comparisons with model inputs without ID. Notations are consistent with Table \ref{tab:main_results}.}
\setlength{\tabcolsep}{0.5em}{
 \resizebox{0.95\textwidth}{!}{
\begin{tabular}{lclcccccccccr}
\toprule
\multicolumn{2}{l}{\makecell{Input Type \& Model $\rightarrow$}} &  & \multicolumn{6}{c}{T} & \multicolumn{3}{c}{T+V} & \multirow{2}{*}{\makecell{\emph{Improv.} \\ w/o ID}} \\
\cmidrule(l){1-2} \cmidrule(l){4-9} \cmidrule(l){10-12}
Dataset & Metric &  &  {\small SASRec} & {\footnotesize Mamba4Rec} & {\small TiM4Rec} & {\small ZESRec} & {\small UniSRec} & {\small VQRec} & {\small MMSRec} & {\small MISSRec} & {\small \modelname{}} & \\
\hline
\midrule
\multirow{4}{*}{Scientific} & R@10 &  & 0.0994 & 0.1118 & 0.1086 & 0.0851 & 0.1188 & 0.1211 & 0.1054 & \underline{\textbf{0.1278}} & \textbf{0.1278} & \cellcolor[HTML]{EFFBEC}- \\
 & R@50 &  &  0.2162 & 0.2149 & 0.2127 & 0.1746 & \underline{0.2394} & 0.2369 & 0.2296 & 0.2375 & \textbf{0.2549} & \cellcolor[HTML]{EFFBEC}6.47\% \\
 & N@10 &  & 0.0561 & 0.0605 & 0.0587 & 0.0475 & 0.0641 & 0.0643 & 0.0548 & \underline{0.0658} & \textbf{0.0668} & \cellcolor[HTML]{EFFBEC}1.52\%  \\
 & N@50 &  &  0.0815 & 0.0829 & 0.0813 & 0.0670 & \underline{0.0903} & 0.0897 & 0.0815 & 0.0893 & \textbf{0.0929} & \cellcolor[HTML]{EFFBEC}2.88\% \\
\midrule
\multirow{4}{*}{Pantry} & R@10 &  & 0.0585 & 0.0586 & 0.0575 & 0.0454 & 0.0636 & 0.0660 & 0.0666 & \underline{0.0771} & \textbf{0.0885} & \cellcolor[HTML]{EFFBEC}14.79\% \\
 & R@50 &  & 0.1647 & 0.1521 & 0.1546 & 0.1141 & 0.1658 & 0.1753 & 0.1801 & \underline{0.1833} & \textbf{0.1878} & \cellcolor[HTML]{EFFBEC}2.45\% \\
 & N@10 &  & 0.0285 & 0.0282 & 0.0287 & 0.0230 & 0.0306 & 0.0293 & 0.0309 & \underline{0.0345} & \textbf{0.0431} & \cellcolor[HTML]{EFFBEC}24.93\% \\
 & N@50 &  & 0.0523 & 0.0484 & 0.0496 & 0.0378 & 0.0527 & 0.0527 & 0.0554 & \underline{0.0571} & \textbf{0.0646} & \cellcolor[HTML]{EFFBEC}13.13\% \\
\midrule
\multirow{4}{*}{Instruments} & R@10 &  & 0.1127 & 0.1170 & 0.1150 & 0.0783 & 0.1189 & \underline{0.1222} & 0.1119 & 0.1201 & \textbf{0.1293} & \cellcolor[HTML]{EFFBEC}5.81\% \\
 & R@50 &  & 0.2104 & 0.2040 & 0.2084 & 0.1387 & 0.2255 & \underline{0.2343} & 0.2219 & 0.2218 & \textbf{0.2426} & \cellcolor[HTML]{EFFBEC}3.54\% \\
 & N@10 &  & 0.0661 & 0.0769 & 0.0741 & 0.0497 & 0.0680 & 0.0758 & 0.0732 & \underline{0.0771} & \textbf{0.0847} & \cellcolor[HTML]{EFFBEC}9.86\% \\
 & N@50 &  & 0.0873 & 0.0988 & 0.0940 & 0.0627 & 0.0912 & \underline{0.1002} & 0.0970 & 0.0988 & \textbf{0.1092} & \cellcolor[HTML]{EFFBEC}8.98\% \\
\midrule
\multirow{4}{*}{Arts} & R@10 &  & 0.0977 & 0.1010 & 0.1026 & 0.0664 & 0.1066 & \underline{0.1189} & 0.1147 & 0.1119 & \textbf{0.1219} & \cellcolor[HTML]{EFFBEC}2.52\% \\
 & R@50 &  & 0.1916 & 0.1939 & 0.1953 & 0.1323 & 0.2049 & \underline{0.2249} & 0.2205 & 0.2100 & \textbf{0.2319} & \cellcolor[HTML]{EFFBEC}3.11\% \\
 & N@10 &  & 0.0562 & 0.0598 & 0.0595 & 0.0375 & 0.0586 & 0.0703 & \underline{0.0719} & 0.0625 & \textbf{0.0739} & \cellcolor[HTML]{EFFBEC}2.78\% \\
 & N@50 &  & 0.0766 & 0.0799 & 0.0796 & 0.0518 & 0.0799 & 0.0935 & \underline{0.0950} & 0.0836 & \textbf{0.0979} & \cellcolor[HTML]{EFFBEC}3.05\% \\
\midrule
\multirow{4}{*}{Office} & R@10 &  &  0.0929 & 0.1075 & 0.1063 &  0.0641 & 0.1013 & \underline{0.1236} & 0.1175 & 0.1038 & \textbf{0.1252} & \cellcolor[HTML]{EFFBEC}1.29\% \\
 & R@50 &  & 0.1580 & 0.1654 & 0.1659 & 0.1113 & 0.1702 & \underline{0.1957} & 0.1859 & 0.1701 & \textbf{0.1999} & \cellcolor[HTML]{EFFBEC}2.15\%\\
 & N@10 &  & 0.0582 & 0.0729 & 0.0708 & 0.0391 & 0.0619 & 0.0814 & \textbf{0.0864} & 0.0666 & \underline{0.0859} & \cellcolor[HTML]{EFFBEC}- \\
 & N@50 &  & 0.0723 & 0.0855 & 0.0837 & 0.0493 & 0.0769 & 0.0972 & \underline{0.1013} & 0.0808 & \textbf{0.1022} & \cellcolor[HTML]{EFFBEC}0.89\% \\
\hline
\bottomrule
\end{tabular}
}
}
\label{tab:main_results_w/o_id}
\end{table*}
\subsection{Comparasion with State-of-the-arts (RQ1)}
\label{sec:RQ1}
The comparative results of model performance are presented in \cref{tab:main_results,tab:main_results_w/o_id}. To ensure a fair comparison, particularly for models like VQRec and MMSRec that do not incorporate ID features, we specifically developed an ID-removed variant of \modelname{} (which eliminates the modality bias described in \S \ref{sec:modality-bias}) to enable equitable performance evaluation under identical conditions. 
\par
Tables \ref{tab:main_results} and \ref{tab:main_results_w/o_id} lead to four main observations. First, textual modality features can effectively supplement or even replace ID features: pretrained text-enhanced models such as UniSRec and VQRec clearly outperform FDSA, S$^3$-Rec, and ZESRec, while our simple text variants of ID-based backbones remain competitive and even surpass their original versions on Pantry and Instruments. Second, under identical settings, Mamba-based models generally outperform Transformer-based counterparts, consistent with the analysis in \S \ref{sec:mamba_greater_than_transformer}. Third, effective transferable pretraining is necessary for strong multi-modal retrieval. Although non-transferable multi-modal models such as M$^3$Rec, HM4SR, and ATHWE improve over many ID-only or text-only baselines, they still lag behind transferable models such as MISSRec and \modelname{}. In particular, HM4SR and ATHWE show that temporal awareness is beneficial, and ATHWE even achieves competitive NDCG on datasets such as Instruments and Office, but without large-scale transferable pretraining multi-modal features are harder to align with recommendation semantics and thus cannot fully realize their retrieval potential. Fourth, \modelname{} achieves the strongest overall performance across most domains, including a 31.78\% NDCG@10 improvement over MISSRec on Pantry. In the ID-removed setting, MISSRec falls behind text-enhanced VQRec, whereas \modelname{} remains superior. Overall, the results indicate that \modelname{} benefits from more complete multi-modal preference modeling, sequence-level semantic alignment, and Mamba's time-aware state-space dynamics. Mechanism analysis of each module is provided in \S \ref{sec:ablation}.
\begin{table}[t]
\centering
\caption{Transfer efficiency}
\setlength{\tabcolsep}{0.6em}{
 \resizebox{0.45\textwidth}{!}{
\begin{tabular}{lclccc}
\toprule
\multicolumn{2}{l}{\makecell{Model $\rightarrow$}} &  & \multirow{2}{*}{MMSRec} & \multirow{2}{*}{MISSRec} & \multirow{2}{*}{MMM4Rec}\\
\cmidrule(l){1-2} 
Dataset & Metric &  & & & \\
\hline
\midrule
\multirow{2}{*}{Scientific} 
 & epochs &  & 25 & 76 & 13 \\
 & s / epoch &  & 2.72 & 2.21 & 2.07 \\
 \midrule
\multirow{2}{*}{Pantry} 
 & epochs &  & 20 & 32 & 10 \\
 & s / epoch &  & 6.43 & 5.97 & 5.58 \\
 \midrule
\multirow{2}{*}{Instruments} 
 & epochs &  & 50 & 65 & 7 \\
 & s / epoch &  & 12.25 & 10.55 & 9.08 \\
 \midrule
\multirow{2}{*}{Arts} 
 & epochs &  & 67 & 166 & 5 \\
 & s / epoch &  & 33.81 & 25.15 & 14.92 \\
 \midrule
\multirow{2}{*}{Office} 
 & epochs &  & 52 & 153 & 5 \\
 & s / epoch &  & 33.57 & 41.06 & 27.93 \\
\hline
\bottomrule
\end{tabular}
}
}
\label{tab:convergence_conparison}
\end{table}
\subsection{Transfer Efficiency (RQ2)}
As shown in \cref{tab:convergence_conparison}, \modelname{} requires fewer fine-tuning epochs and less time per epoch than MMSRec and MISSRec, particularly on Arts and Office. We attribute this efficiency to shared-weight alignment without contrastive objectives and SR-aligned time-aware masking without clustering. Together with the RQ1 results, these findings show that \modelname{} improves both retrieval quality and transfer efficiency.

\subsection{Model Analyses (RQ3)}
\label{sec:ablation}
We design six variants to study the contribution of each key component:
\begin{itemize}[leftmargin=5pt]
    \item (1) w/o PT: Trained directly on downstream datasets without pretraining. 
    \item (2) w/o Time: Removes time-aware enhancement components from TiSSD and TiCoSSD.
    \item (3) w/o Shared: Eliminates cross-modal TiSSD shared-weight constraints during multi-modal alignment. 
    \item (4) w/o LF: Removes the Learnable Filter, i.e., $\widehat{\mathcal{D}}^{f} = \mathcal{F}^{-1}\left(\widetilde{\mathcal{D}}^{v} + \widetilde{\mathcal{D}}^{t}\right)$.
    \item (5) w/o AF: Removes the Adaptive Filter while retaining the Learnable Filter, i.e., $\widehat{\mathcal{D}}^{f} = \mathcal{F}^{-1}\left(\widetilde{\delta}\left(\widetilde{\mathcal{D}}^{v} + \widetilde{\mathcal{D}}^{t}\right)\right)$.
    \item (6) 2L: Stacking 2-layer TiSSD and TiCoSSD. 
\end{itemize}\begin{table}[h] %
\caption{Ablation study.}
        \setlength{\tabcolsep}{0.2em}{
	\resizebox{\columnwidth}{!}{
        \begin{tabular}{lcccccccc}
        \toprule
         & \multicolumn{4}{c}{Scientific} & \multicolumn{4}{c}{Office} \\
        \cmidrule(l){2-5} \cmidrule(l){6-9}
        Variant & R@10 & R@50 & N@10 & N@50 & R@10 & R@50 & N@10 & N@50 \\
        \hline
        \midrule
        (0) \modelname{} & \textbf{0.1348} & \textbf{0.2627} & \textbf{0.0724} & \textbf{0.1002} & \underline{0.1337} & \underline{0.2132} & \underline{0.0906} & \underline{0.1080} \\
        \cmidrule(l){1-1}
        (1) \phantom{-} w/o PT & 0.1257 & 0.2399 & 0.0647 & 0.0896 & 0.1178 & 0.1868 & 0.0751 & 0.0901 \\
        (2) \phantom{-} w/o Time & \underline{0.1328} & \underline{0.2559} & 0.0696 & 0.0965 & 0.1329 & 0.2128 & 0.0895 & 0.1069 \\ 
        (3) \phantom{-} w/o Shared & 0.1294 & 0.2521 & 0.0685 & 0.0955 & 0.1331 & 0.2124 & 0.0897 & 0.1069 \\ 
        (4) \phantom{-} w/o LF & 0.1303 & 0.2592 & 0.0685 & 0.0968 & 0.1314 & 0.2090 & 0.0886 & 0.1056 \\
        (5) \phantom{-} w/o AF & 0.1317 & 0.2517 & 0.0687 & 0.0949 & 0.1310 & 0.2083 & 0.0891 & 0.1060 \\
        \midrule
        (6) \phantom{-} 2L & 0.1309 & 0.2544 & \underline{0.0698} & \underline{0.0969} & \textbf{0.1343} & \textbf{0.2140} & \textbf{0.0926} & \textbf{0.1101} \\
        \bottomrule
        \end{tabular}
	}
        }
	\label{tab:ablation_study}
\end{table}
As shown in \cref{tab:ablation_study}, variant (1) validates pretraining, variant (2) validates time-aware enhancement, and variant (3) validates cross-modal TiSSD weight sharing. Variants (4) and (5) confirm the necessity of the two-stage frequency fusion design, since removing either filter degrades performance. Variant (6) further suggests that deeper backbones may overfit on Scientific but improve results on the larger Office dataset.

\subsection{Additional Experimental Analyses}
Due to space limitations, detailed efficiency analyses, experiments on full-modality subsets, and extended module analyses are provided in the GitHub repository linked in the abstract.

\section{Conclusions}
\modelname{} addresses inefficient transfer in multi-modal sequential recommendation through algebraic constraints for sequence-aware alignment and fusion. It combines time-aware state-space decay, cross-modal weight sharing, and sequence-level multi-modal fusion within a unified cross-entropy training framework. Experiments show superior retrieval performance, faster downstream convergence, and robustness in ID-removed and modality-missing settings, indicating that SR-compliant algebraic constraints can jointly support multi-modal effectiveness and transfer efficiency.

\begin{acks}
We thank the anonymous reviewers and the meta-reviewer for their valuable comments, which were instrumental in improving the quality of this work. We also thank all individuals who contributed to this work.
\end{acks}



\bibliographystyle{ACM-Reference-Format}
\bibliography{quote}

\appendix

\section{Related Work}
\subsection{Sequential Recommendation}
The field of Sequential Recommendation (SR) has evolved from traditional Markov chain-based \cite{FPMC2010} approaches to contemporary deep learning paradigms. Early deep architectures encompassed CNN-based models (e.g., Caser \cite{Caser2018}), RNN-based designs (e.g., GRU4Rec \cite{GRU4Rec2016}), and Transformer-driven frameworks (e.g., SASRec \cite{SASRec2018}). While Transformer \cite{Transformer2017}-based models achieved superior performance in complex interaction scenarios through their powerful attention mechanisms, their quadratic complexity relative to sequence length prompted the development of efficient alternatives. Subsequent architectures like MLP-based FMLP-Rec \cite{FMLPRec2023} and LRU \cite{LRU2023}-based LRURec \cite{LRURec2024} sought to balance computational efficiency with recommendation accuracy.
Recent advancements leverage architectural inductive biases aligned with SR characteristics. Mamba4Rec \cite{Mamba4Rec2024} exemplifies this trend, where the Mamba \cite{Mamba2023} architecture's inherent sequence modeling priors enable both efficiency and performance gains, particularly in long interaction sequences. Building upon State Space Duality (SSD) \cite{SSD2024} developments in structured state space models (SSM), next-generation frameworks like TiM4Rec \cite{TiM4Rec2024} further advance SR through temporal-aware enhancements, achieving new Pareto frontiers in the accuracy-efficiency trade-off.
\par
Note that all the aforementioned models are based on pure ID feature modeling. As discussed in the introduction, such approaches face significant limitations in recommendation performance and knowledge transfer. Researchers have gradually introduced additional information to enrich item representations and enhance model capabilities: FDSA \cite{FDSA2019} and $S^{3}$-Rec \cite{S3-Rec2020} improve ID backbone performance by integrating pre-extracted textual features into IDs; MM-Rec \cite{MM-Rec2022} employs VL-BERT \cite{VL-BERT2020} for fused image-text representation learning; CARCA \cite{CARCA2020} incorporates multi-modal features into item embeddings via cross-attention mechanisms; MMMLP \cite{MMMLP2023} successfully adapts the MLP-Mixer \cite{MLP-Mixer2021} architecture to SR. $\text{M}^3$Rec \cite{M3Rec2025} integrates MoE architecture, pioneering the application of Mamba to multi-modal SR. While these works partially address the shortcomings of pure ID-based modeling, they remain suboptimal in achieving universal multi-modal sequential representations and effective transfer learning capabilities.

\subsection{Pre-training and Transfer Learning in Recommendation}
Since raw multi-modal information cannot be directly utilized in recommendation semantic spaces, acquiring sufficient prior knowledge through large-scale pre-training to transform multi-modal features into recommendation-oriented semantics becomes critical for enhancing multi-modal sequential recommendation (MMSR) performance. Though conceptually similar to cross-domain recommendation, the "pre-train and transfer" paradigm offers greater flexibility by eliminating the need for cross-domain correspondences over overlapping items.
Existing approaches diverge in transfer strategies: user-centric methods like RecGURU \cite{RecGURU} employ adversarial learning to improve generalized user representations across domains, while more effective item-centric approaches focus on multi-modal utilization. For instance, ZESRec \cite{ZESRec2021} directly adopts pre-extracted text embeddings as transferable item representations, UniSRec \cite{UniSRec2022} learns transferable text semantics via parameter whitening techniques, and VQRec \cite{VQRec2023} enhances UniSRec's transferability through vector quantization.
\par
Introducing visual modalities (beyond text) significantly increases modeling complexity due to cross-modal alignment challenges between visual and textual modalities. While works \cite{MMSRec2023, PMMRec2024} like MMSRec \cite{MMSRec2023} address this via computationally intensive self-supervised contrastive learning strategies, such manually crafted constraints often degrade convergence speed, particularly during fine-tuning on new domains. Although MISSRec \cite{MISSRec2023} balances performance and transfer efficiency through dynamic candidate-side fusion and parameter-efficient tuning, its multi-modal interest aggregation method, designed to filter redundant information (inherently addressing contribution imbalance), compromises end-to-end learning via suboptimal heuristic filtering, ultimately limiting fine-tuning convergence.
Our MMM4Rec advances the pre-train-transfer paradigm with two key innovations:
(\romannumeral1) By designing model-inherent algebraic constraints that encompass two-stage algebraic constraints for multi-modal alignment and fusion aligned with the SR principle, we eliminate complex optimization objectives and procedures, achieving effective modeling through a simple consistent cross-entropy loss in both pre-training and fine-tuning phases, thus enabling transfer-efficient multi-modal sequential recommendation.
(\romannumeral2) By leveraging state space decay properties of State Space Duality and specialized time-aware constraints, we resolve the uneven item information contribution problem in MMSR without resorting to suboptimal manual feature engineering (e.g., clustering methods in MISSRec). This framework enables rapid capture of critical item information in user interaction sequences, achieving breakthroughs in fine-tuning convergence efficiency and multi-modal retrieval performance.

\section{Additional Efficiency and Fairness Analysis}
\label{sec:additional_efficiency}
This section complements the transfer-efficiency results in the main paper with a comprehensive comparison of computational cost, a detailed description of the controlled evaluation protocol, and complete fine-tuning curves on all five downstream domains.

\subsection{Comprehensive Computational Efficiency}
\cref{tab:comprehensive_efficiency} compares the model size, FLOPs, memory consumption, and wall-clock efficiency of the three methods. In addition to the forward cost of encoding a user sequence, we separately measure the cost of full-corpus candidate retrieval because this stage can dominate computation when the target-domain item set is large.

\begin{table*}[t]
\centering
\caption{Comprehensive efficiency comparison on the five downstream domains. Each entry follows the order MMSRec / MISSRec / \modelname{}. FLOPs are measured per user instance; Fwd., Ret., and Total denote sequence forward encoding, full-corpus candidate retrieval, and their sum, respectively. GPU memory is the peak allocated memory during fine-tuning. Train and Infer. denote the wall-clock time per training and evaluation epoch, while Conv. is the total fine-tuning time to the best early-stopped checkpoint. Lower values are better, and the best result in each entry is highlighted in bold.}
\label{tab:comprehensive_efficiency}
\setlength{\tabcolsep}{2.7pt}
\resizebox{\textwidth}{!}{%
\begin{tabular}{lcccccccc}
\toprule
Dataset & \makecell{Param. $\downarrow$\\(M)} & \makecell{Fwd. FLOPs $\downarrow$\\(M)} & \makecell{Ret. FLOPs $\downarrow$\\(M)} & \makecell{Total FLOPs $\downarrow$\\(M)} & \makecell{GPU Memory $\downarrow$\\(GB)} & \makecell{Train Time $\downarrow$\\(s)} & \makecell{Conv. Time $\downarrow$\\(s)} & \makecell{Infer. Time $\downarrow$\\(s)} \\
\hline
\midrule
Scientific  & 9.78 / 7.53 / \textbf{4.24}   & 371.2 / 146.8 / \textbf{105.2} & \textbf{4.49} / 7.89 / 7.86    & 375.7 / 154.7 / \textbf{113.1} & 4.63 / 3.29 / \textbf{3.01} & 2.72 / 2.21 / \textbf{2.07}  & 68.0 / 168.0 / \textbf{26.9}    & 0.32 / 0.44 / \textbf{0.19} \\
Pantry      & 10.30 / 8.21 / \textbf{4.51}  & 382.3 / 160.4 / \textbf{105.2} & \textbf{5.02} / 8.82 / 8.78    & 387.4 / 169.2 / \textbf{114.0} & 4.61 / 3.55 / \textbf{3.06} & 6.43 / 5.97 / \textbf{5.58}  & 128.6 / 191.0 / \textbf{55.8}   & 0.41 / 0.67 / \textbf{0.30} \\
Instruments & 15.50 / 14.91 / \textbf{7.10} & 453.4 / 185.0 / \textbf{105.2} & \textbf{10.20} / 17.94 / 17.86 & 463.6 / 203.0 / \textbf{123.1} & 5.72 / 4.47 / \textbf{3.11} & 12.25 / 10.55 / \textbf{9.08} & 612.5 / 685.8 / \textbf{63.6}   & 0.89 / 1.35 / \textbf{0.58} \\
Arts        & 26.84 / 29.55 / \textbf{12.76}& 650.8 / 224.3 / \textbf{105.2} & \textbf{21.52} / 37.84 / 37.68 & 672.3 / 262.1 / \textbf{142.9} & 8.27 / 5.31 / \textbf{3.72} & 33.81 / 25.15 / \textbf{14.92}& 2265.3 / 4174.9 / \textbf{74.6} & 1.37 / 2.64 / \textbf{1.05} \\
Office      & 22.40 / 36.13 / \textbf{15.30}& 556.9 / 234.8 / \textbf{105.2} & \textbf{17.09} / 46.78 / 46.58 & 574.0 / 281.6 / \textbf{151.8} & 7.42 / 5.42 / \textbf{3.91} & 33.57 / 41.06 / \textbf{27.93}& 1745.6 / 6282.2 / \textbf{139.7}& 2.54 / 5.12 / \textbf{2.01} \\
\hline
\bottomrule
\end{tabular}%
}
\end{table*}

Compared with MISSRec, \modelname{} reduces total per-instance FLOPs by 26.9--46.1\% and uses 43.7--57.7\% fewer parameters across the five domains. Its peak GPU memory is also reduced by 8.5--30.4\%, corresponding to an average per-domain reduction of 22.1\%. The reduction becomes more pronounced on the larger domains, showing that the efficiency advantage is not limited to small-scale evaluation settings.

The decomposition of FLOPs further explains this behavior. The forward cost of \modelname{} remains constant at 105.2M FLOPs across domains because multi-modal alignment and fusion are performed on the user-sequence side. By contrast, the forward costs of MMSRec and MISSRec grow with the target-domain item scale because they involve candidate-side multi-modal modeling. MMSRec obtains lower retrieval FLOPs by pooling candidate-side textual and visual features before scoring, whereas MISSRec and \modelname{} compute modality-specific candidate scores and then aggregate them. Nevertheless, the substantially lower sequence-modeling cost of \modelname{} yields the lowest total FLOPs, training time, convergence time, and inference time on every domain.

\subsection{Fair Comparison Protocol}
To ensure that the efficiency results reflect architectural and optimization differences rather than mismatched experimental conditions, all three transferable multi-modal recommenders use the same pre-training domains and downstream data splits. We retain the optimizer configurations and validated learning rates recommended by the corresponding methods; the selected learning rates are all on the order of $10^{-4}$. Fine-tuning uses a batch size of 1024 and early stopping with a patience of 10 epochs for every model. All efficiency measurements are reproduced with the same data pipeline, CUDA/PyTorch software environment, and a single NVIDIA RTX 4090D GPU. Parameters, FLOPs, peak memory, and wall-clock times are therefore measured under a controlled and directly comparable setting.
\begin{figure}[ht]
    \centering
    \includegraphics[width=0.9\columnwidth]{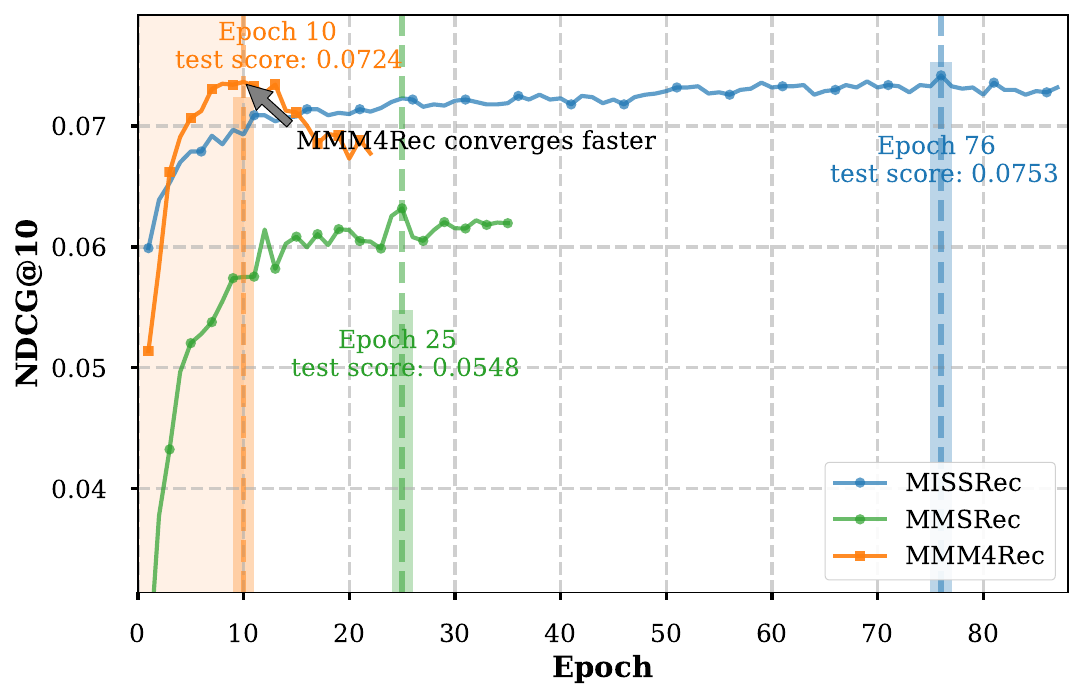}
    \caption{Comparison of model convergence speed on Scientific.}
    \label{fig:convergence_scientific_appendix}
\end{figure}
\begin{figure}[ht]
    \centering
    \includegraphics[width=0.9\columnwidth]{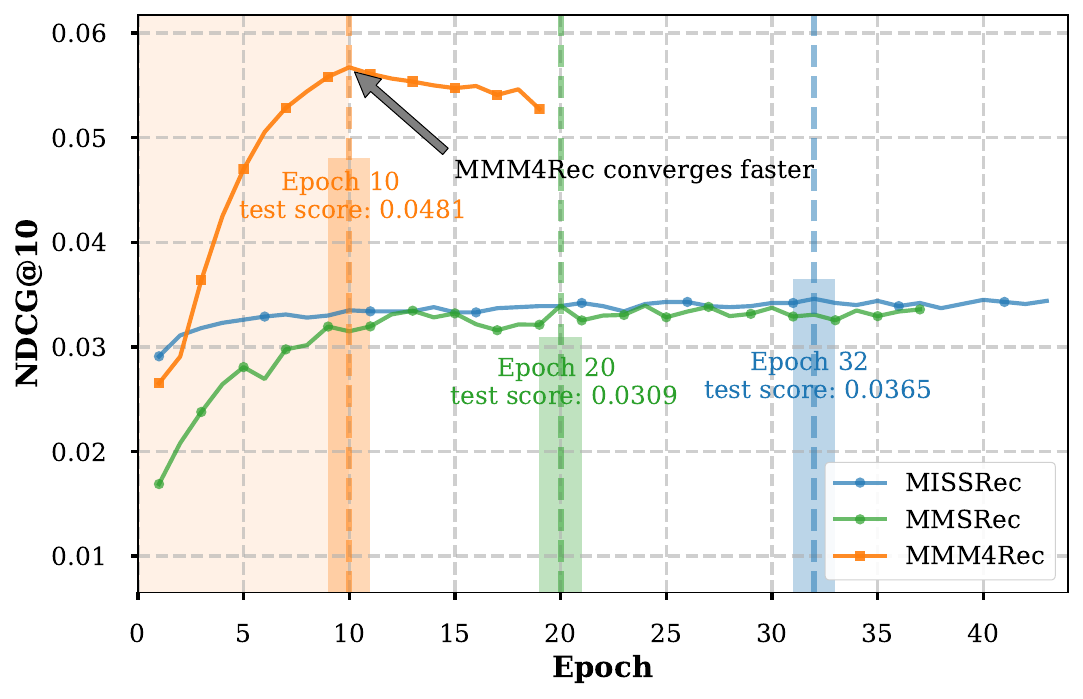}
    \caption{Comparison of model convergence speed on Pantry.}
    \label{fig:convergence_pantry_appendix}
\end{figure}
\begin{figure}[ht]
    \centering
    \includegraphics[width=0.9\columnwidth]{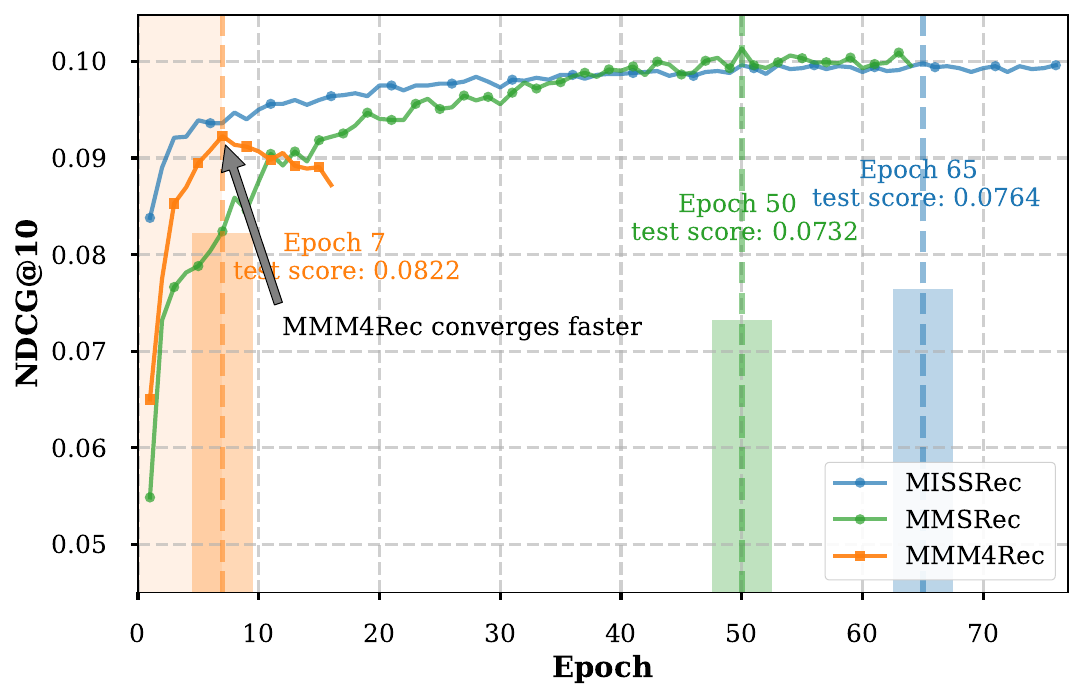}
    \caption{Comparison of model convergence speed on Instruments.}
    \label{fig:convergence_instruments_appendix}
\end{figure}
\begin{figure}[ht]
    \centering
    \includegraphics[width=0.9\columnwidth]{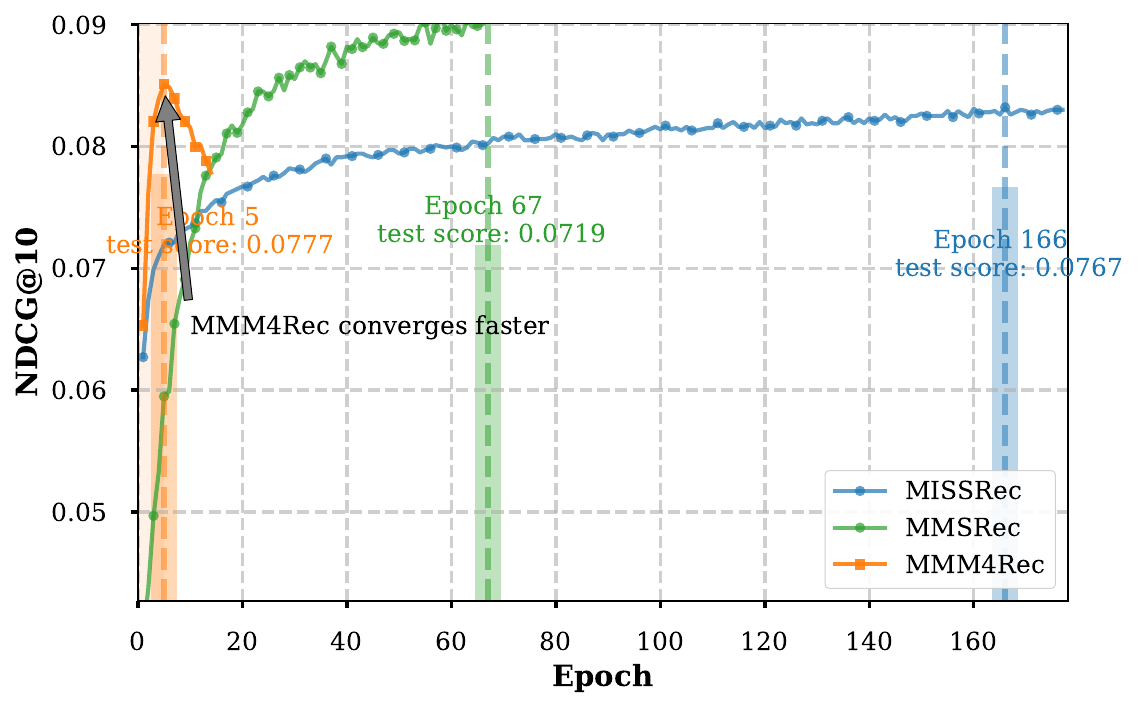}
    \caption{Comparison of model convergence speed on Arts.}
    \label{fig:convergence_arts_appendix}
\end{figure}
\begin{figure}[ht]
    \centering
    \includegraphics[width=0.9\columnwidth]{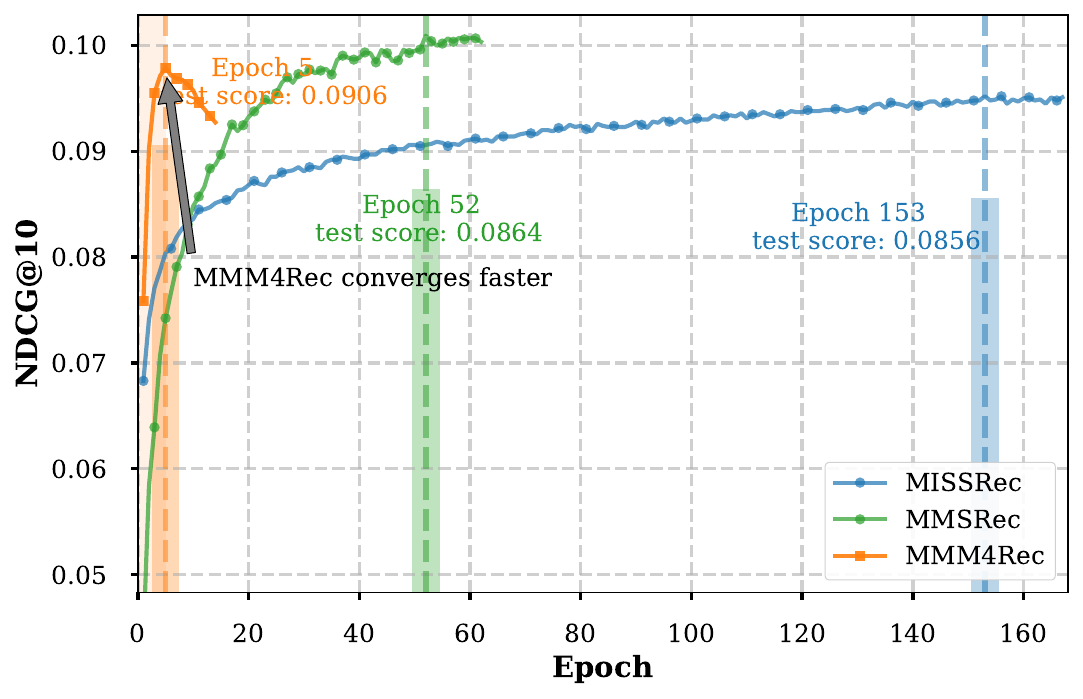}
    \caption{Comparison of model convergence speed on Office.}
    \label{fig:convergence_office_appendix}
\end{figure}
\subsection{Convergence Behavior}
To complement the aggregate efficiency measurements, we report the complete fine-tuning curves on all five downstream datasets in \cref{fig:convergence_scientific_appendix,fig:convergence_pantry_appendix,fig:convergence_instruments_appendix,fig:convergence_arts_appendix,fig:convergence_office_appendix}. These curves show that \modelname{} reaches its best validation performance with fewer fine-tuning epochs than MMSRec and MISSRec across all target domains. The difference is particularly substantial on Arts and Office, where \modelname{} reaches a stable solution within only a few epochs.

The convergence results and the wall-clock measurements in \cref{tab:comprehensive_efficiency} provide complementary evidence. Weight-shared sequence-level alignment avoids the additional contrastive objectives used by MMSRec, while the time-aware state-space masks directly encode temporal contribution patterns without the clustering-based aggregation procedure of MISSRec. Together with the unified cross-entropy objectives used in pre-training and fine-tuning, these properties make downstream adaptation both computationally lighter and easier to optimize.

\section{Additional Ablation with Attention-based Replacements}
\label{sec:attention_replacement}
To examine whether the gains of TiSSD and TiCoSSD arise merely from their projection and fusion structures, we construct a controlled attention-based variant of \modelname{}. Specifically, standard self-attention replaces TiSSD in the sequence-level alignment stage, and cross-attention replaces TiCoSSD in the multi-modal fusion stage. The corresponding query/key/value-style projections and the surrounding alignment and fusion pipeline are retained; the principal change is that the state-space decay and time-aware masks are replaced by attention masks.

\begin{table}[ht]
\centering
\caption{Comparison between \modelname{} and its attention-based replacement on four downstream domains. The replacement uses standard self-attention in place of TiSSD and cross-attention in place of TiCoSSD, while retaining the remaining projection and fusion pipeline.}
\label{tab:attention_replacement}
\setlength{\tabcolsep}{3.4pt}
\resizebox{\columnwidth}{!}{%
\begin{tabular}{llcccc}
\toprule
Dataset & Variant & R@10 & R@50 & N@10 & N@50 \\
\hline
\midrule
\multirow{2}{*}{Scientific}
 & \modelname{} & \textbf{0.1348} & \textbf{0.2627} & \textbf{0.0724} & \textbf{0.1002} \\
 & Attention replacement & 0.1310 & 0.2503 & 0.0695 & 0.0955 \\
\midrule
\multirow{2}{*}{Office}
 & \modelname{} & \textbf{0.1337} & \textbf{0.2132} & \textbf{0.0906} & \textbf{0.1080} \\
 & Attention replacement & 0.1236 & 0.1959 & 0.0851 & 0.1009 \\
\midrule
\multirow{2}{*}{Instruments}
 & \modelname{} & \textbf{0.1330} & \textbf{0.2525} & \textbf{0.0822} & \textbf{0.1082} \\
 & Attention replacement & 0.1209 & 0.2233 & 0.0751 & 0.0976 \\
\midrule
\multirow{2}{*}{Arts}
 & \modelname{} & \textbf{0.1307} & \textbf{0.2486} & \textbf{0.0777} & \textbf{0.1034} \\
 & Attention replacement & 0.1109 & 0.2062 & 0.0662 & 0.0869 \\
\hline
\bottomrule
\end{tabular}%
}
\end{table}

As shown in \cref{tab:attention_replacement}, the original model consistently outperforms the attention-based replacement for all 16 dataset--metric combinations. The performance gaps are particularly clear on Instruments and Arts; for example, replacing the two state-space modules decreases R@50 from 0.2525 to 0.2233 on Instruments and from 0.2486 to 0.2062 on Arts. Because the controlled variant retains the projection and fusion pathway, this consistent degradation cannot be explained simply by additional module capacity. Instead, it supports the importance of the SR-specific inductive biases encoded by TiSSD and TiCoSSD: state-space decay prioritizes recent interactions, the time-aware correction preserves temporally distant but informative behaviors, and the cross-modal state-space mask introduces the same temporal structure into multi-modal fusion. These results complement the time-aware ablation in the main paper and show that standard attention is not an equivalent substitute for the proposed temporal constraints.

\section{Additional Full-modality Results}
To further validate \modelname{} under complete image availability, we report an additional comparison on the full-modality subset of the Office domain in \cref{tab:full-modal}. Removing items with missing image modalities yields larger gains for \modelname{}, which is consistent with the main-text observation that better visual coverage further benefits transferable multi-modal retrieval.

\begin{table}[h]
\centering
\caption{Comparisons on the full-modality Office subset}
\setlength{\tabcolsep}{0.2em}{
 \resizebox{\columnwidth}{!}{
\begin{tabular}{lclcccccr}
\toprule
\multicolumn{2}{l}{\makecell{Model $\rightarrow$}} &  & \multirow{2}{*}{UniSRec} & \multirow{2}{*}{MMSRec} & \multirow{2}{*}{MISSRec} & \multirow{2}{*}{ATHWE} & \multirow{2}{*}{MMM4Rec} & \multirow{2}{*}{\emph{Improv.}}\\
\cmidrule(l){1-2} 
Dataset & Metric &  & & & & \\
\hline
\midrule
\multirow{4}{*}{Office} 
 & R@10 &  & 0.1407 & 0.1344 & \underline{0.1421} & 0.1323 & \textbf{0.1467} & \cellcolor[HTML]{E9EAFF}3.24\%\\
 & R@50 &  & 0.2203 & 0.2105 & \underline{0.2223} & 0.1991 & \textbf{0.2237} & \cellcolor[HTML]{E9EAFF}0.63\%\\
 & N@10 &  & 0.0957 & 0.0969 & 0.0966 & \underline{0.1009} & \textbf{0.1080} & \cellcolor[HTML]{E9EAFF}7.04\%\\
 & N@50 &  & 0.1133 & 0.1146 & 0.1138 & \underline{0.1155} & \textbf{0.1249} & \cellcolor[HTML]{E9EAFF}8.14\%\\
\hline
\bottomrule

\end{tabular}
}
}
\label{tab:full-modal}
\end{table}

\end{document}